# INVESTIGATION OF NON-STABLE PROCESSES IN CLOSE BINARY RY SCUTI

M.I. KUMSIASHVILl, R.Sh. NATSVLISHV1L1, K.B. CHARGEISHV1LI
E. Kharadze National Astrophysical Observatory of Chavchavadze State University
Kazbegi ave.2a, 0160, Tbilisi, Georgia
kumsiashvili@genao.org, rezonats@yahoo.com, ketichargeishvili@yahoo.com

We present results of reanalysis of old electrophotometric data of early type close binary system RY Scuti obtained at the Abastumani Astrophysical Observatory, Georgia, during 1972-1990 years and at the Maidanak Observatory, Uzbekistan, during 1979-1991 years. It is revealed non-stable processes in RY Sct from period to period, from month to month and from year to year. This variation consists from the hundredths up to the tenths of a magnitude. Furthermore, periodical changes in the system's light are displayed near the first maximum on timescales of a few years. That is of great interest with regard to some similar variations seen in luminous blue variable (LBV) stars. This also could be closely related to the question of why RY Sct ejected its nebula.

***1. Introduction.*** RY Scuti is a unique massive binary star system in a rare transitional evolutionary phase. It is surrounded by a young circumstellar nebula. The system is thought to be a rare progenitor of a WR+OB system, so it may be a "Rosetta Stone" for understanding the late evolutionary stages of close binary stars - particularly the formation of close WR binaries by tidally-induced mass transfer and mass loss. So, the close binary RY Sct undergoes mass-transfer and mass exchange processes and has complicated structure in its outer atmospheric layers. It possibly suffers evolutionary changes in a comparatively short time interval that is significant for constructing an evolutionary model of the system.

The short history of investigation of this binary system is as follows: Merrill [1] found the He II λ4686 line in the spectrum of RY Sct being specific only to nebulae and their nuclei. There are strong forbidden lines of [Fe III] and [Si III] in the spectrum, as well as emission lines of hydrogen, helium and other elements. RY Sct was observed as a radio source [2]. The presence of the intense emission lines and radio emission (as in β Lyr) favored the assumption that there may be a small H II region around RY Sct [3] . The star is strongly reddened by the surrounding gas and dust shell [4]. Based on spectroscopic observations, Cowley and Hutchings [5] found that the binary system consists of two early-type supergiants with a mass ratio of $q=M_2/M_1=1.25$. Guirichin and Mardirossian [6] suggested that the secondary must be surrounded by a geometrically thick accretion disk, which can explain its anomalously low luminosity, and that RY Sct is similar to the peculiar β Lyr system and is currently on its way to become a Wolf-Rayet (WR) system. Milano et al. [7] and Guirichin and Mardirossian [6] analyzed the photometric light curves of Ciatti et al. [8] assuming a component mass ratio of $q=M_2/M_1=1.25$ by Cowley and Hutchings [5]. Later a number of authors working within the framework of an International coordinated program (the Abastumani Astrophysical Observatory was the coordinator of this program) used the spectroscopic data obtained by R. West to infer a new component mass ratio $q=M_2/M_1=3.3$ [9]. Spectral observations of RY Sct have been carried out by R. West in the region λλ3450-5160Å with a dispersion of 12Å /mm.

Antokhina and Cherepashchuk [10] adopted this new component mass ratio and applied the synthetic light curve method to interpret photometric observations made by: Kumsiashvili [11], Ciatti et al. [8] and Zakirov [12] of RY Sct in the V band. The model, which incorporates the secondary component in the form of a geometrically thick disk, satisfactorily describes the light curves and is consistent with the observed features of the binary. It was concluded that the parameters of RY Sct make it look very much like a WR+OB binary if we assume that the less massive star is at the end of the stage of preliminary mass transfer and is now exposing its helium core on its way to become a WR system.

By Sahade et al.[13] RY Sct was reinvestigated on the basis of spectroscopic material, obtained at the European Southern Observatory (ESO) at La Silla, at the Cerro Tololo Inter-American Observatory (CTIO), covering the spectral regions 3400-5150Å and 5700-6700Å, and with IUE. In that paper, they concluded that RY Sct is an interacting binary formed by a BO primary of -10 $M_{\odot}$ and a secondary of -36 $M_{\odot}$. Among its emission features, there is a set of lines that are characteristic of planetary nebulae. The secondary component of the system, which is surrounded by an opaque envelope, emits in the He II λ4686.

A "new era" of investigation of RY Sct began with articles by the American group of authors based on the analysis of HST observations and high-resolution ground-based data in multiple spectral regions. The results mainly concern the physical characteristics of the compact nebula around RY Sct. They have obtained and analyzed rich high-quality observational material in various spectral intervals: visual, IR and radio. They had spectral material obtained at ESO, HST, KPNO, and CTIO. In addition, they had HST, Palomar 5m, Keck 10 m telescope images, and Very Large Array radio continuum maps. On the basis of this material they published detailed analyses of the peculiar nebula around RY Sct [14,15,16,17]. From HST and Keck images, they first showed directly that around the close binary there exists a double-lobed 1"x2" nebula, corresponding to radii of a few thousand AU at a distance of 1.8 kpc. The American Investigators determined masses of the companion stars (49 and 39 $M_{\odot}$) and a total bolometric luminosity of 3.4 x $10^6$ $L_{\odot}$ [18]. This value is near the Eddington limit for this system. For spectral classes of the companion stars they received O 9.5 and O 6.5 and a distance between them - 0.43 AU.

They estimated also the systemic velocity -20±3 km/s; the electron density $n_e$=2x $10^5$ $cm^3$ and the temperature in the nebula of RY Sct - $9000^oK<T<10000^oK$ [16]. The expansion rate of the nebula at the southwest side is twice larger than that at the northeast side. For the mass of the nebula they obtained M≈0.003 $M_{\odot}$.

Smith et al. [15] studied the proper motions of the outer double-torus nebula and they conclude the gas in this HII region was probably ejected from the central binary around the year 1876±20.

The observation that the nebula appears to have been ejected recently (~120 years ago) during an outburst of the star [15] combined with strong asymmetry in the nebula, may suggest that this short-lived evolutionary phase is characterized by sporadic mass loss. Available evidence [15] implies that RY Scuti may have suffered something analogous to the S Doradus outbursts of luminous blue variables (LBVs), although this type of outburst might appear phenomenological different in a contact binary system.

The kinds of gas outflows expected in systems like RY Scuti are explored in a series of recent papers by Nazarenko and Glazunova [19, 20, 21]. They present hydrodynamical simulations in two and three-dimensions to model the case of interacting binary β Lyr. RY Scuti and β Lyr share many of properties common to the W Serpentis class. Therefore, the massive binaries that are just emerging from the rapid mass transfer phase probably belong to the observed class of W Serpentis binaries [22].

The recent spectroscopic study of the RY Sct system [23] supported the fact that the radius of the massive companion is significantly less than its critical Roche radius and therefore ruled out the over contact model. They find the masses of the components are $M_1=7.1\pm1.2M_{\odot}$ and $M_2=30.0\pm2.1M_{\odot}$. Their mass results are quite similar to these first obtained by Skul′skii [24] but are lower than these determined by Sahade et al. [13]. They estimate the mass ratio is $M_2/M_1=4.2\pm0.7$. Moreover this study strongly suggests the existence an accretion disc surrounding the smaller more massive star. This model fits the observations very well and can explain some photometric behavior of RY Sct [25].

After these articles the Georgian authors of this article decided to return to observational material of an old project on RY Sct, in which the Abastumani Astrophysical Observatory was a coordinator.

For revealing RY Sct phenomenon it is essential to investigate what kind of non-stationary processes occur in this very peculiar close binary system.

***2.Old photoelectric observations – new results.*** During 1972-1985 periods UBV electrophotometric observations were made in Abastumani for 103 nights [11].

As mentioned above, following the American investigators' data one component could experience the ejection to 2 magnitudes resulting, in the formation of a nebulous the system. Due to such a sporadic ejection some nonstable process, of comparatively less intensity, are probable going .on in the system. Such a phenomenon is specific to a certain type of LBV variables. Just therefore, in terms of above mention three-color UBV photoelectric data comprising the interval of 14 years, the average value of each night for phases and brightnesses was calculated. The results are presented in table 1. The phase dependent light curve in V color was plotted also (Fig. 1).

Table.1.

| JD | phase | ΔV | ΔB | ΔU |
|---|---|---|---|---|
| 2441510.3490 | 0.2771 | -0.332 | 0.614 | -0.017 |
| 2441537.2950 | 0.6993 | -0.364 | 0.556 | -0.071 |
| 2441858.3535 | 0.5586 | -0.187 | 0.807 | 0.019 |
| 2441888.3548 | 0.2554 | -0.359 | 0.621 | -0.074 |
| 2441890.3391 | 0.4337 | 0.123 | 1.087 | 0.423 |
| 2442221.3711 | 0.1896 | -0.303 | 0.674 | -0.056 |
| 2442244.3204 | 0.2524 | -0.322 | 0.610 | -0.055 |
| 2442248.3350 | 0.6133 | -0.317 | 0.693 | -0.001 |
| 2442250.3243 | 0.7922 | -0.288 | 0.681 | -0.028 |
| 2442273.2790 | 0.8555 | -0.093 | 0.852 | 0.224 |
| 2442653.2397 | 0.0094 | 0.152 | 1.091 | 0.418 |
| 2442663.2893 | 0.9128 | 0.250 | 1.211 | 0.516 |
| 2442930.4396 | 0.9265 | 0.263 | 1.252 | 0.616 |
| 2442933.4123 | 0.1936 | -0.252 | 0.700 | 0.062 |
| 2442955.3739 | 0.1677 | -0.210 | 0.740 | 0.102 |
| 2442960.3647 | 0.6163 | -0.218 | 0.732 | 0.091 |
| 2442982.2936 | 0.5875 | -0.179 | 0.778 | 0.168 |
| 2442984.2844 | 0.7664 | -0.301 | 0.644 | -0.022 |
| 2442986.3359 | 0.9509 | 0.271 | 1.245 | 0.639 |
| 2442990.3361 | 0.3104 | -0.236 | 1.739 | 0.044 |
| 2443332.3060 | 0.0495 | -0.020 | 0.949 | 0.298 |
| 2443341.3338 | 0.8609 | -0.051 | 0.925 | 0.315 |
| 2443342.3053 | 0.9483 | 0.254 | 1.258 | 0.670 |
| 2443346.3116 | 0.3084 | -0.283 | 1.688 | 0.038 |
| 2443391.2611 | 0.3488 | -0.138 | 1.853 | 0.107 |
| 2443420.2048 | 0.9505 | 0.290 | 1.258 | 0.651 |
| 2443423.2018 | 0.2199 | -0.285 | 1.679 | 0.026 |
| 2443424.2070 | 0.3103 | -0.255 | 0.694 | 0.054 |
| 2443425.2084 | 0.4003 | 0.004 | 0.959 | 0.298 |
| 2443429.1963 | 0.7587 | -0.331 | 0.625 | -0.051 |
| 2443430.1923 | 0.8482 | -0.103 | 0.861 | 0.200 |
| 2443666.4215 | 0.0826 | -0.136 | 0.825 | 0.167 |
| 2443670.3752 | 0.4379 | 0.142 | 1.111 | 0.463 |
| 2443671.4100 | 0.5309 | -0.115 | 0.878 | 0.185 |
| 2443672.4284 | 0.6224 | -0.244 | 0.727 | 0.104 |
| 2443673.4549 | 0.7147 | -0.315 | 0.651 | 0.012 |
| 2443687.3650 | 0.9650 | 0.273 | 1.261 | 0.648 |
| 2443688.3752 | 0.0558 | -0.045 | 0.915 | 0.282 |
| 2443695.3920 | 0.6866 | -0.307 | 0.652 | -0.004 |
| 2443696.4057 | 0.7777 | -0.305 | 0.678 | 0.028 |
| 2443697.4103 | 0.8680 | 0.002 | 0.977 | 0.383 |
| 2443698.4255 | 0.9592 | 0.268 | 1.250 | 0.620 |
| 2443720.3100 | 0.9264 | 0.219 | 1.190 | 0.540 |
| 2443727.3059 | 0.5553 | -0.138 | 0.825 | 0.113 |
| 2443730.3289 | 0.8270 | -0.188 | 0.750 | 0.093 |

Table.1.

| JD | phase | $\Delta V$ | $\Delta B$ | $\Delta U$ |
|---|---|---|---|---|
| 2443731.3279 | 0.9169 | 0.166 | 1.153 | 0.459 |
| 2443751.2525 | 0.7079 | -0.285 | 0.652 | 0.022 |
| 2443779.2218 | 0.2219 | -0.297 | 0.649 | -0.004 |
| 2443781.2386 | 0.4032 | 0.028 | 0.976 | 0.318 |
| 2443782.2556 | 0.7979 | -0.265 | 0.670 | 0.053 |
| 2443783.2818 | 0.5868 | -0.197 | 1.173 | 0.041 |
| 2443787.2214 | 0.9409 | 0.238 | 1.203 | 0.516 |
| 2444049.3174 | 0.5003 | 0.029 | 0.996 | 0.378 |
| 2444055.3434 | 0.0419 | -0.006 | 0.951 | 0.299 |
| 2444086.3550 | 0.8295 | -0.201 | 0.752 | 0.104 |
| 2444434.3334 | 0.1086 | -0.150 | 0.829 | 0.138 |
| 2444437.3565 | 0.3803 | -0.026 | 0.945 | 0.240 |
| 2444438.3251 | 0.4674 | 0.225 | 1.187 | 0.468 |
| 2444442.4018 | 0.8339 | -0.116 | 0.863 | 0.116 |
| 2444459.2731 | 0.3504 | -0.156 | 0.787 | 0.167 |
| 2444494.2263 | 0.4923 | 0.030 | 0.985 | 0.336 |
| 2444754.4205 | 0.8807 | 0.033 | 1.006 | 0.379 |
| 2444765.4452 | 0.8716 | -0.035 | 0.947 | 0.264 |
| 2444809.3172 | 0.8152 | -0.216 | 0.753 | 0.088 |
| 2444810.3309 | 0.9064 | 0.174 | 1.148 | 0.525 |
| 2444811.3327 | 0.9967 | 0.172 | 1.155 | 0.525 |
| 2444814.2917 | 0.2624 | -0.317 | 0.624 | -0.035 |
| 2444870.2578 | 0.2930 | -0.337 | 0.631 | -0.091 |
| 2444871.2515 | 0.3824 | -0.090 | 0.863 | 0.215 |
| 2444872.2691 | 0.4738 | 0.105 | 1.061 | 0.412 |
| 2445116.4321 | 0.4212 | 0.084 | 1.045 | 0.341 |
| 2445117.4006 | 0.5083 | -0.004 | 0.952 | 0.305 |
| 2445118.3972 | 0.5978 | -0.244 | 0.723 | 0.045 |
| 2445133.3278 | 0.9399 | 0.243 | 1.221 | 0.535 |
| 2445135.3606 | 0.1227 | -0.254 | 1.720 | 0.032 |
| 2445137.3725 | 0.3035 | -0.276 | 0.667 | -0.011 |
| 2445140.3873 | 0.5745 | -0.171 | 0.784 | 0.126 |
| 2445144.3731 | 0.9328 | 0.196 | 1.189 | 0.501 |
| 2445145.3567 | 0.0212 | 0.093 | 1.062 | 0.409 |
| 2445150.4511 | 0.4786 | 0.075 | 1.026 | 0.336 |
| 2445176.3382 | 0.8061 | -0.257 | 0.707 | 0.039 |
| 2445196.3239 | 0.6025 | -0.262 | 0.690 | 0.001 |
| 2445198.2832 | 0.7786 | -0.293 | 0.658 | 0.012 |
| 2445199.3188 | 0.8680 | -0.041 | 0.922 | 0.213 |
| 2445231.2333 | 0.7404 | -0.323 | 0.647 | -0.023 |
| 2445232.2171 | 0.8289 | -0.208 | 0.756 | 0.100 |
| 2445259.1830 | 0.2526 | -0.361 | 0.625 | 0.079 |
| 2445260.1665 | 0.3412 | -0.228 | 0.745 | 0.070 |
| 2445527.3405 | 0.3571 | -0.189 | 0.768 | 0.156 |
| 2445528.3640 | 0.4490 | 0.115 | 1.065 | 0.401 |
| 2445579.2388 | 0.0220 | 0.055 | 1.019 | 0.359 |
| 2445616.2009 | 0.3446 | -0.117 | 0.835 | 0.172 |
| 2445617.1879 | 0.4332 | 0.114 | 1.093 | 0.447 |
| 2445875.3652 | 0.6403 | -0.239 | 0.748 | 0.081 |
| 2445886.3855 | 0.6309 | -0.226 | 0.760 | 0.065 |
| 2445915.3188 | 0.2316 | -0.214 | 0.762 | 0.074 |
| 2445968.2290 | 0.9877 | 0.167 | 1.159 | 0.499 |
| 2446226.4232 | 0.1966 | -0.270 | 0.677 | -0.016 |
| 2446227.4357 | 0.2873 | -0.226 | 0.734 | 0.009 |
| 2446238.3715 | 0.2705 | -0.316 | 0.668 | -0.003 |
| 2446258.3554 | 0.0666 | -0.096 | 0.862 | 0.164 |
| 2446286.2675 | 0.5757 | -0.208 | 0.738 | 0.113 |
| 2446317.2796 | 0.3632 | -0.201 | 0.782 | 0.045 |

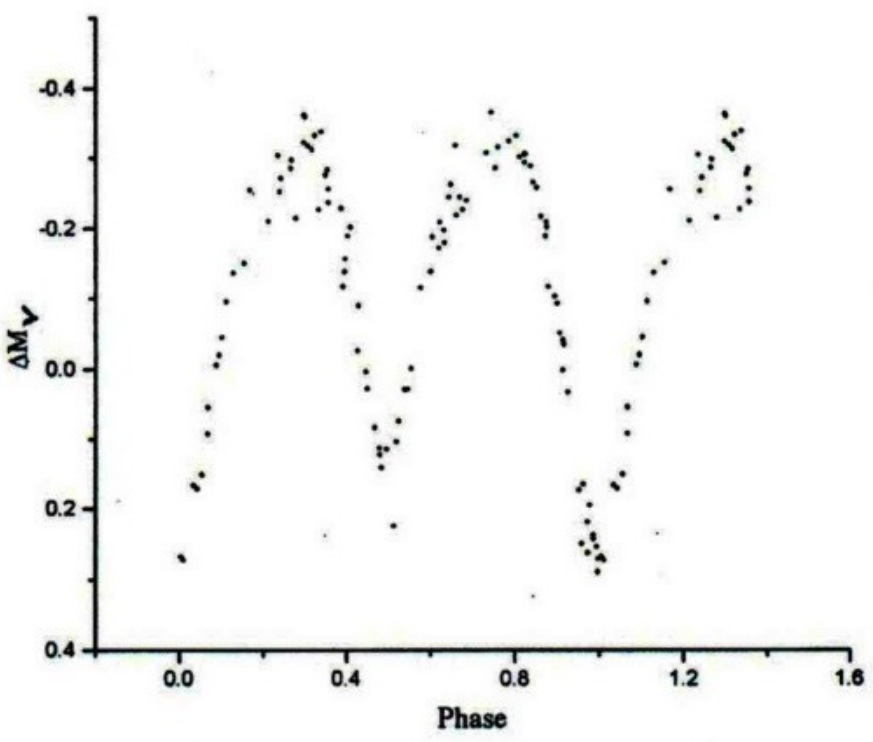

Fig. 1. V mean light curve of RY Scuti (1972-1985 years)

Analysis showed that during the same phase in different years the light varies from the hundredth up to the tenth of a magnitude. For instance: if to begin from the light minimum it is noticeable that from June 12-13 till 23-24, 1982 the eclipse intensity grew by $0^m.05$. This means that the variability occurred during a period. It is evident that the eclipse depth varies by $0^m.12$ year after year. Accordingly, a conclusion can be drawn that the eclipse depth varies during several days, months and years as well.

The picture is similar at the phase maximum. In particular, on august 2-3, 1984 the light decrease by about $0^m.08$ at 0.277 phase is observed compared with other years. The same event is noticed for 1985: at about the same phase (the same maximum phase) from 10-11 till 21-22 June the stellar light changed by approximately $0^m.08$. This variability also occurred during a period.

The light variation near 0.4 phase is essential on the curve. In particular, the star is brighter at 0.402 phase than at 0.390 by about $0^m.07$; I.e. instead of the decrease with the phase increase towards the secondary minimum the rise in the light is observed. The same happens at 0.425 and 0.427 phases. The star is brighter near the last phase by about $0^m.065$, while the contrary must be the case. The phase of 0.512 calls attention. Here is July 17-18, 1980 observation. The depth of the secondary minimum rose by $0^m.1$ compared with other years. Comparison of observations performed in different years shows that the depth similarly varies at 0.5 phase by about $0^m.05$. The light fluctuations are observed at the secondary maximum as well, at about 0.75 phase making $0^m.08$.

Also six-color photoelectric observations of RY Set were performed in Abastumani with the cooperation program [27]. The star was observed in the Stromgren medium-band system (ubvy) and $H_\beta$ line in 1983-1990 during 49 nights with 1.25 m telescope. Two hundred and thirty individual measurements were done in each filter. For revealing non-stable processes from period to period, from month to month or year after year a mean value for each night was estimated in six colors. The v color curve is plotted in Figure 2. Small-scale micro-variations are clearly seen on the curve. In more details it can be said that the light periodic variation of $0^m.05$ amplitude can possibly occur at the same phase year after year. For instance: if in 1983 the value was equal to $1^m.041$ at 0.165 phase in v, in 1984 at 0.167 phase it falls dawn to $1^m.089$ and in 1986 it was $1^m.035$ at 0.168 phase, i.e. it almost returned to the value observed nearly at the same phase in 1983. The observations at 0.252, 0.255, 0.256 and 0.265 phases carried out in 1983, 1986, 1984 and 1990 respectively are noteworthy. The derived values $0^m.985$, $0^m.972$, $0^m.982$, $0^m.972$ are coincident with the accuracy of $0^m.01$, but the value obtained in 1983 at 0.273 makes $0^m.877$ being different from $0^m.985$ at 0.252 phase by $0^m.1$. This means that a two very close phases in the same year (June 17-18, August 12-13, 1983) the star brightened by $0^m.1$ and later on it returned to its initial value.

A situation similar to three-color observations is observed in six color ones, when the following phase of the descending branch shows more brightening. For instance $1^m.016$ is consistent with 0.323 in v, while at phase 0.324 the value is $0^m.969$, being by $0^m.046$ brighter compared with the first one. The first value is suitable to observations of 21-22.06.1985 and the other one corresponds to those of 29-30.08.1986. In 1983-1989 the fluctuations in the secondary minimum is about $0^m.04$. It should be noted that the stellar brightness at the following two phases 0.569 and 0.571 are $1^m.002$ and $1^m.183$ and correspond to them respectively. The difference in brightnesses is about $0^m.18$. Besides, our observation on 13-14.07.1983 and that on 26-27.08.1989 are consistent with the first and second phases. This means, that during the mentioned night of 1989 the star brightened by about $0^m.2$ on the ascending branch of the secondary minimum. Year after year (1983-1989) fluctuations are likewise observed in the secondary maximum being of the order of about $0^m.07$. Peculiarities on the descending branch of the primary minimum are noteworthy. Since that time when the brightness was $1^m.025$ at 0.803 phase on 10-11.09.1985, the star brightened up to $0^m.997$ at the following 0.840 phase on 29-30.08.1989. Than at 0.866 on 27-28.06.1985 the bright falls down to $1^m.087$, i.e. year after year fluctuations make $0^m.1$ at the beginning of the descending branch. There is a case when the light on the descending branch is exactly the same at 0.870 phase in different years. For instance on 16-17.08.1985 and 24-25.08.1986 the brightness was $l^m.095$.

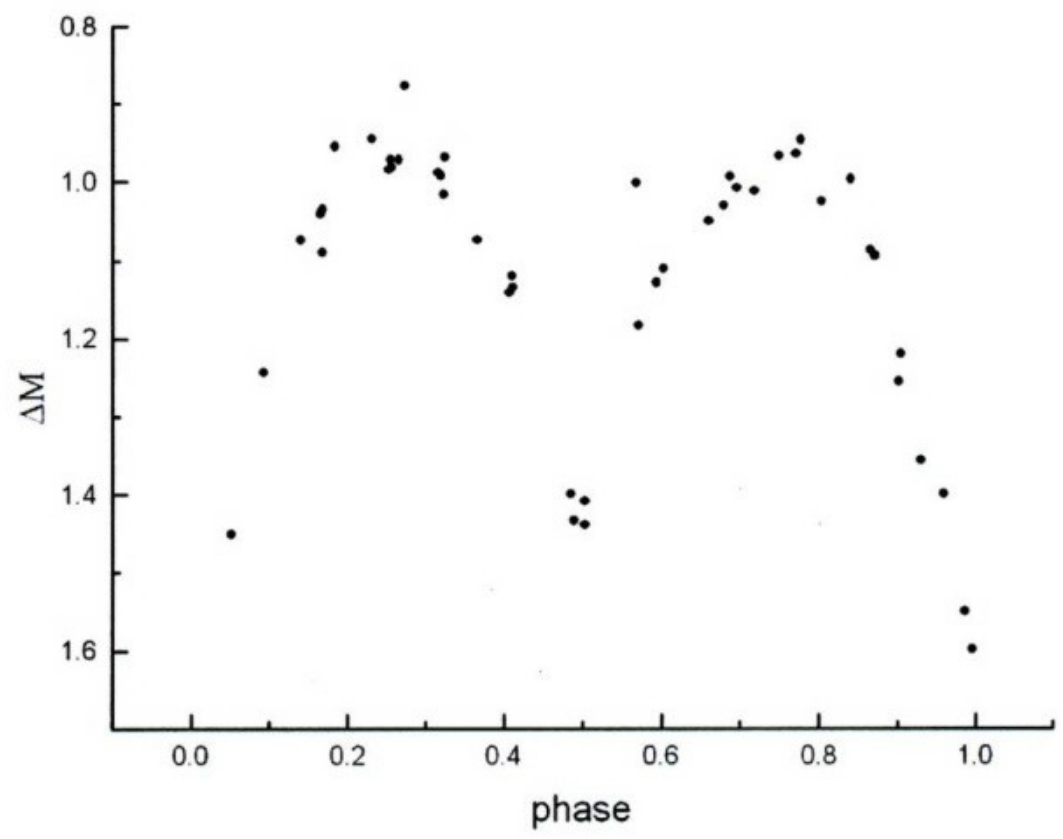

Fig.2. v light curve of Stromgren bands.

Within the same cooperation program five-color UBVRI photoelectric observations were also performed on Mt. Maidanak in 1979-1981 during 100 nights by Zakirov [12]. The star recommended by M. Kumsiashvili was used as the comparison star. Certain asymmetry in the primary minimum is observed at the first and last contacts; the descending branch is steeper than part of the curve emerging from the eclipse. Evidently such an asymmetry of the primary minimum could be explained by the effect of gas flows in the system. Peculiarity of the secondary minimum consists in the curve bend at 0.565 phase of the ascending branch. Due to lacking of the observation data Zakirov failed to observe the same bending at 0.435 phase symmetric to 0.5. However, the behavior of the upper and lower parts of the curve favors such a bending. At the same time it should be noted that this assumption is confirmed by the behavior of the curve, resulted from three-color observations, on the descending branch near the indicated phase. Fluctuations, coming up to $0^m.07$, were likewise seen on Zakirov's light variation curve. Although we could not succeed in searching for a fast variability of this variable during a night at different phases.

In term of the above observations it can be concluded that the light values at the same phase differing by a period are unlike and this difference clearly exceeds the observational precision. Accordingly, there may be a cyclic variability of the amplitude of the light non-stable fluctuations. A search for this event is of great interest.

In our opinion, these results favor the Americans' assumption that one of the components is probably LBV type variable. They think that the star experienced 2 magnitude equivalent outburst 120 years ago, specific to this type variables. In consequence a relatively less intensity microvariation is observed in this system. Any periodicity of this variability has not been revealed yet.

***3. periodical variations in light.*** Analysis of our three-color, six-color and Zakirov's five-color observations it is obvious that sometimes the variable returns to its initial state. This was especially seen in the first maximum of the curve. Therefore we took the plunge, based on our three-color and Zakirov's five-color observations, to begin in V color with plotting the night mean values according to JDs for the first and second maxima to be observed.

Kumsiashvili's observations were turned into magnitudes at $\varphi$=0.170-0.355 and $\varphi$=0.640-0.850 phases, i.e. at the first and second maxima moments. The derived night mean values together with Zakirov's observations were plotted (Fig.3, Fig. 4) according years (1972-1991) for both maxima separately. It turned out on the curve appropriate to the first maximum certain periodicity of variability is observed, whereas the same cannot be said about the second maximum. According to the time this variability takes place in the following way: approximately every four years the curve falls down to minimum; then in the same period it increases to maximum and finally it decreases to minimum, i.e. from minimum to minimum this period makes about 8 years. The same is confirmed with observations till 1991 found by us later.

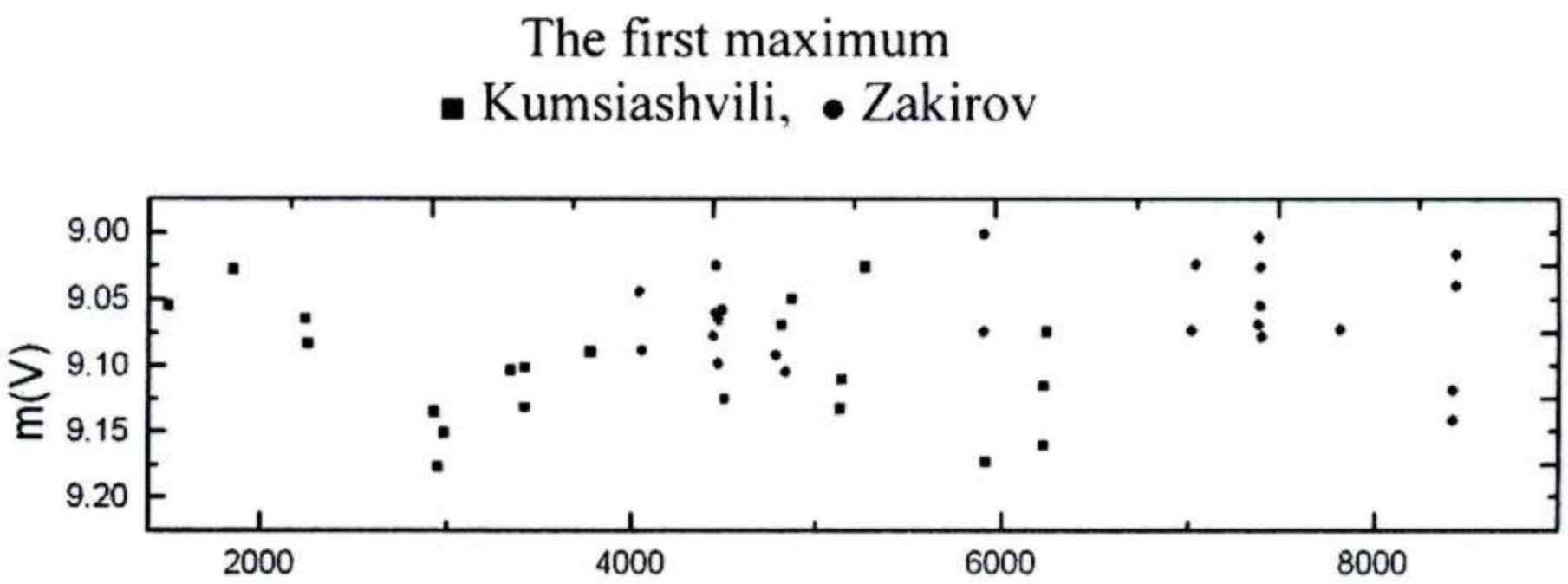

Fig.3. J.D.(2440000+)

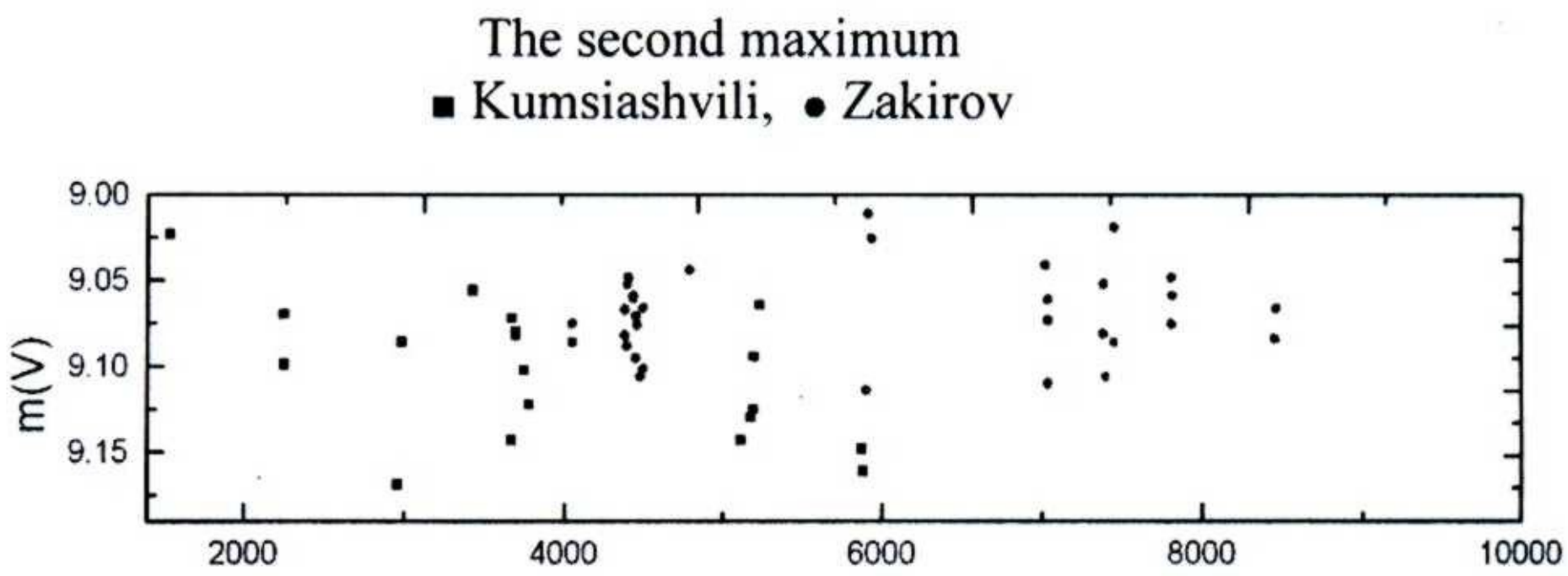

Fig.4. J.D.(2440000+)

As it is seen RY Sct photometry has never been excessive. At every observation, in future, a new phenomenon in the light variability of the system can be revealed.

***4. About period variation.*** An assumption on the sporadic mass eject is assumed to exist. Besides, there are plenty of photometric and spectral data on non-stable processes taking place in them. Resent extra-atmospheric, infrared and radio observations point to the above mention also. Naturally, an idea arose: to examine the problem of period variations based on the photometric data at our disposal. The more so that there are certain grounds after the earlier photographic data had been reviewed. For instance, in the framework of the collaboration program, following our request, Belserene [28] revised the Harvard Observatory (USA) plate library containing RY Sct observations since 1888. Having analyzed the data from the standpoint of period variations the scientist concluded that there are small parabolic variations.

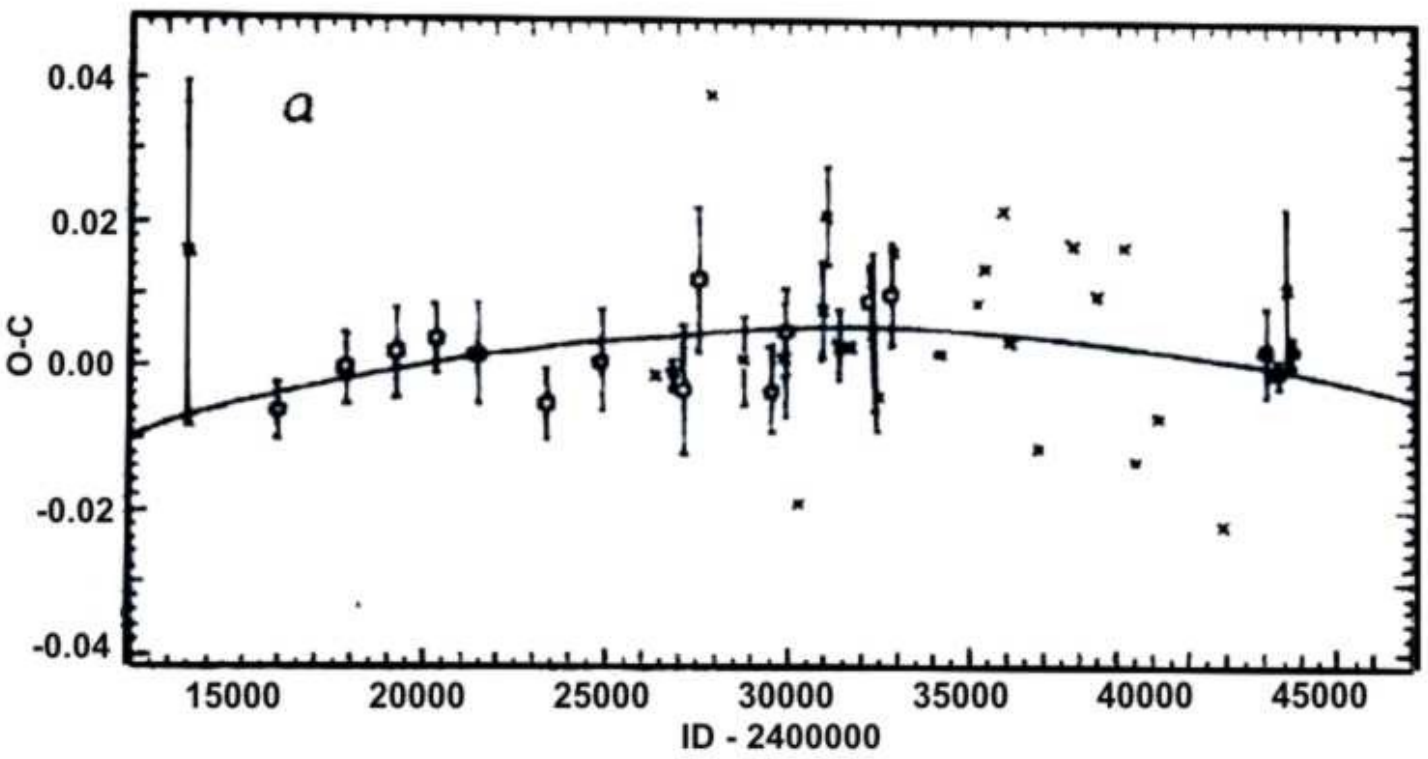

Fig. 5.

Variation of O-C deviation against the Julian day is presented (Fig. 5). O-C is the difference between the observed and estimated minima. They are presented in the figure together with those of Italian researchers [7] confirming the above variation. Crosses denote these data.

The plate library of Moscow Sternberg Astronomical Institute (observations of 1899-1978) was revised likewise under the supervision of Cherepashchuk [29]. To derive the calculated minimum and data to be comparable, the same Belserene elements were used: Min=JD2443342.456+11$^d$.1246E

E is the number of all the periods elapsed from the initial minimum. Cherepashchuk has presented all the data accessible in references. Basically the photographic data are concerned. According to the plot (E-defendent O-C) he infers that there is no reason for a serious period variation of RY Scuti.

Being the period variation, based on the Abastumani photoelectric material, not studied yet we decided to carry out countings in this direction. At first the observed minima moments were fixed. A mean curve of the primary minima and then the individual ones observed at minimum were plotted. Using the same Belserene's elements, the theoretical minima, suitable to our individual ones, differences of observed and estimated minima, i.e. O-C values and our E-ones were drawn on the Cherepashchuk [29] curve (Fig. 6).

Here crosses denote Kumsiashvili' s data. As for the period variation, this phenomenon is not pronounced. We think that for finally solving the problem is necessary to carry out long series of precise photoelectric observations exactly in the depth of the primary minimum.

**5. New spectral and polarimetric observation.** Spectral observations of RY Sct were made using 2.6 m telescope of the Byurakan Astrophysical Observatory (Armenian National Academy of Sciences) in August 2005. The telescope is equipped with CCD device "SCORPIO". The size of CCD is 2058x2063 pixels. Cooling was made by liquid nitrogen. Dispersion is 1.7 Å/pix and spectral region is 4000-7150Å. Observations were made during 3 nights 5, 9 and 12 August of 2005. 24 spectrograms were obtained. There were revealed such kind rapid spectral changes, which do not depend on the orbital phase of the system. For example, August 12 there was fixed strong emission line $H_\delta$ and this line was not on the previous and next spectrograms. By our spectral observations, the spectral region near $H_\delta$ line is variable and these changes are due to nearest spectral lines of other elements (He I; S II).

Evidently it is reasonable that we have changes, which do not depend on the orbital phase. The old spectral observations (1981) we have also show inner activity of the system components. In 1200-3000 Å diapason for some moment, which do not depends on the phase, were fixed spectral lines $\lambda$1525.31, $\lambda$1911.5 and $\lambda$2195. These are emission lines for some moment and they are not seen for other moments. The $\lambda$2195 line is often emission but sometimes it is absorption.

In our new data also it is seen changes, which do not depend on phase for continuum spectrum. Because this, it is advisable spectral monitoring of RY Sct at any time for as long spectral region as possible.

Polarimetric observations of RY Sct were made at the Abastumani Astrophysical Observatory by 1.25 m telescope during 9 nights - 8-9.06.2005; 1011.06.2005; 3-4.07.2005; 27-28.07.2005; 29-30.07.2005; 31.07-01.08.2005; 67.08.2005; 8-9.08.2005; 6-7.09.2005. Observations were made in the integral light. Collecting time was 18 seconds. Automatic Scanning Electropolarimeter (ASEP-78) was made in Abastumani Astrophysical Observatory. It is equipped with different light-filters but we could not observe the system with the filters. It is principally significant polarimetric observations in different irradiative diapasons as polarization degree often depends on the wavelength of radiation. This dependence is significant for the concretization of physical parameters of polarizing environment. For example, for Thomson scattering of radiation the degree of polarization does not depend on the wavelength but in the case of Relay scattering it depends on the wavelength.

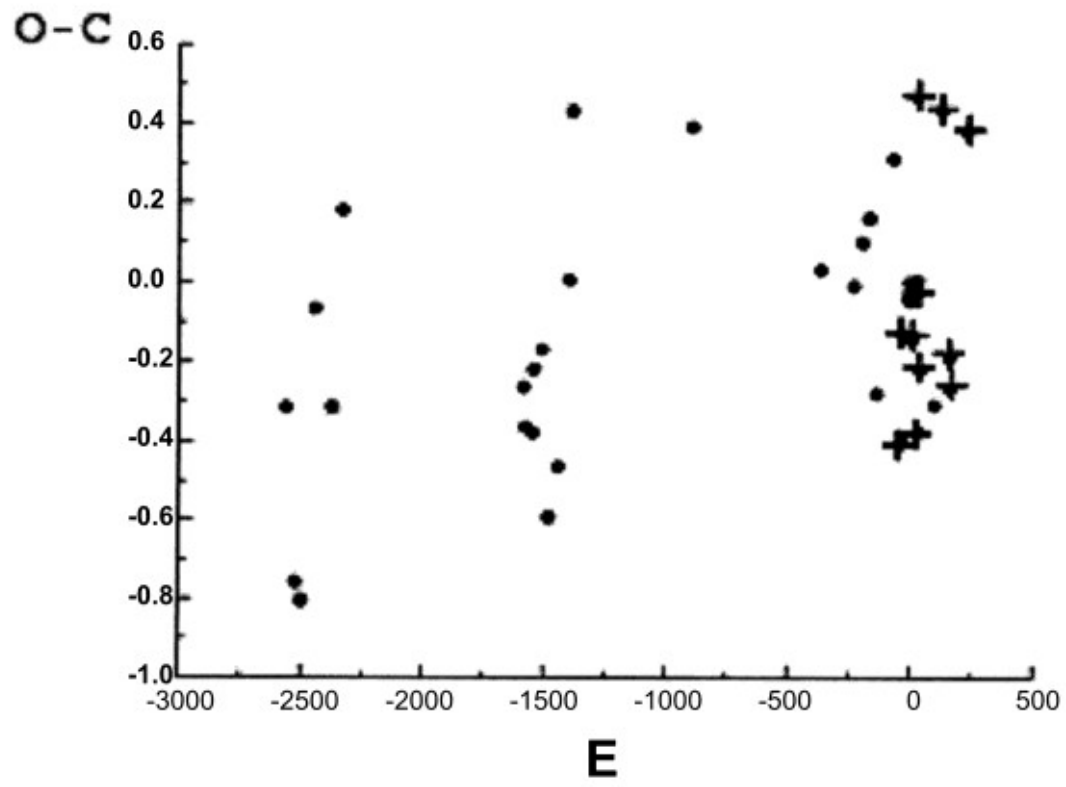

Fig. 6.

It is observed the very fast changes of integral light polarization degree. But the obtained material is insufficient for to consider them as reliable. After fixing rapid spectral changes, we think that it is possible rapid changes of proper polarization degree of light too. Polarization position angle changed in certain range (Table 2).

Table.2.

| Date | UT | phase | P% | $\theta^O$ |
|---|---|---|---|---|
| June 8,2005 | 22" 44$^m$ | 0.80 | 2.78 | 32.4 |
| June 10, 2005 | 22 26 | 0.97 | 2.58 | 34.6 |
| July 3, 2005 | 21 41 | 0.04 | 2.56 | 31.2 |
| July 27, 2005 | 20 57 | 0.19 | 2.81 | 29.8 |
| July 29, 2005 | 19 54 | 0.37 | 2.84 | 30.5 |
| July 31,2005 | 21 16 | 0.55 | 2.68 | 28.9 |
| August 6, 2005 | 20 42 | 0.09 | 2.72 | 32.4 |
| August 8, 2005 | 20 26 | 0.27 | 2.80 | 29.5 |
| September 6, 2005 | 19 07 | 0.87 | 2.54 | 28.5 |

Parameters of polarization are averaged for each date. Time and phases are given for mean moment of an observation.

Really RY Sct system locates in the Galaxy plane ($b \approx 0^o$) and on the distance of 1.8 kps, so it may be rather good amount of interstellar matter at its direction. But because of the observed structure around this object we can suppose that the large quantity of gas and dust is near this system.

If scattering layers around the system have stable state, the magnetic field would be cause of this and the polarization mechanism of electromagnetic waves may be the Thompson scattering of radiation on free electrons or the scattering on asymmetric aligned dust particles. In other case these layers must be lost in space because of the light pressure and stellar wind. So, certain structure of the matter around the system revealed by American colleagues shows the existence of the magnetic field.

We think that it is rather difficult to explain complicated structure of the P Cyg profile of the emission line He I 3888.6, in the conditions of spherical-symmetrical envelope, even if it includes several layers. By considering the widths of absorption parts of $H_\gamma$ and He I 4471.5 lines we can suppose that the two-component absorption part of He I 3888.6 in fact represents an absorption line with inner emission. And in such a case it is impossible to represent its profile by the common envelope's spherical-symmetrical layers. We think that the contour of this line arises from the environment of non-spherical symmetry, which consists of fragments and layers moving against the line of sight. Throw out of dens expanding fragments may be rapid casual processes not depending on the orbital phase, but induced by components inner activity. The structure of the top of emission formation of absorption part of the He I 3888.6 line would be depend on the motion of these fragments against the line of sight. As if the environments between two different cone surfaces with common top in the center of the system creates the emission formation in the wide absorption part of the line He I 3888.6.

So, on the base of the existing and new observational data, we think that the observational properties of RY Sct directly connect on the state of the system's components and surrounding environment - on its structure and dynamics. But the geometry of the system is a result of those inner physical processes, which are characterize for components on recent evolutionary stage. So, observations show great number of rapid changes in the system RY Sct, which are results of inner processes taking place in the components on certain evolution phase.

***6. Conclutions.***

1. According to Kumsiashvili's three-color UBV photoelectric data comprising the interval of 14 years and Zakirov's UBVRI ones the average value of each night for phases and brightnesses was calculated. The phase dependent V light curve was plot. Analysis showed that during the same phase in different years the light varies from the hundredth up to the tenth of a magnitude. It is evident that the eclipse depth varies by $0^m.12$ year after year. Accordingly, a conclusion can be drawn that the eclipse depth varies from period to period, from month to month and year after year.
2. Similarly of above mentioned, it is revealed certain periodical light changes in the first maximum but the same cannot be said about the second maximum. This result favor the assumption of American colleagues that one of the components is probably LBV type variable.
3. Not revealed period changes according available photometric material. But for final solving the problem it is very significant the special precise photometric observations in the depts of the primary minimum during a time as long as possible.

We thank our Uzbek colleague, Prof. M. Zakirov for his observations and Armenian colleagues A. Karapetian, G.Ohanian, S. Balayan and T.Movsesian for helping in spectral observations. Support was provided by CRDF through Bilateral Grants Program (project GEPI-3333-TB-03).

## ***References***

1. P.W. Merrill, Astrophys.J., **67**, 179.1928.
2. V.A.Hughes, A.W. Woodsworth, Nature Phys. Sci., **242**, 116, 1973.
3. A.R.King,R.F. Jameson, A&A, **71**, 326. 1979.
4. G.L. Grasdalen,J.A. Hackwell, R.D. Gehrz, D. McClain, Asrophys. J., **234**, L129,1979.
5. A.P, Cowley, J.B. Hutchings, PASP, **88**, 456, 1976.
6. G. Guirichin, F. Mardirossian, A&A, **101**, 138, 1981.

7. L. Milano, A. Vittone, F. Ciatti, A. Mammano, R.Margoni, R., G.Strazzulla, A&A, **100**, 59, 1981
8. F. Ciatti, A.*Mammano, R.* Margoni, L. Milano, G. Strazzulla, A. Vittone, Astron. Asrophys. Suppl, **41**, 143, 1980.
9. M.Yu. Skul'skii, Bull. Abastumani Astrophys. Obs., **58**, 101,1985.
10. E.A. Antokhina, A.M. Cherepashchuk, AZh Pis'ma, **14**, 252, 1988.
11. M.I. Kumsiashvili, Bull. Abastumani Astrophys. Obs., **58**, 61, 1985.
12. M.M. Zakirov, Bull. Abastumani Astrophys. Obs., **58,** 425, 1985.
13. J. Sahade, RW. West and M.Yu. Skul'skii, Rev. Mexicana Astron.Astrofis., **274**, 259, 2002.
14. N. Smith, RD. Gehrz, R.M. Humphreys *et al.,* Astron.J., **118**, 960, 1999.
15. N. Smith, RD. Gehrz, W.M. Goss, Astron.J., **122**, 2700, 2001.
16. N. Smith, RD. Gehrz, O. Stahl *et al.,* Astrophys.J., **578,** 464, 2002
17. RD. Gehrz, N. Smith, B. Jones *et al.,* Astrophys.J., **559**, 35, 2001.
18. RD. Gehrz, T.L. Hayward, J.R Houck *et al.,* Astrophys.J., **439**, 417, 1995.
19. V.V.Nazarenko, L.V.Glazunova, Astr.Rep., **47**, 1013, 2003.
20. V.V.Nazarenko, L.V.Glazunova, Astr.Rep., **50**, 369,2006a.
21. V.V.Nazarenko, L.V.Glazunova, Astr.Rep., **50,** 380,2006b.
22. A.E.Tarasov, ASP Conf.Ser., **214**,644, 2000.
23. E.D. Grundstrom, et.al., Astrophys.J., **667**, 505, 2007.
24. M.Yu. Skul'skii, Soviet Ast., **36,** 411, 1992.
25. G. Djurasevich, I. Vince, O. Atanackovich, Astron.J., **136**, 767, 2008.
26. M.l. Kumsiashvili, Bull. Abastumani Astrophys. Obs., **58**, 93, 1985.
27. E.A. Antokhina, M.l. Kumsiashvili, Astron. Zh. Pis'ma, **25,** 662, 1999.
28. E.P. Belserene, private communication (1982).
29. A.M. Cherepashchuk, Bull. Abastumani Astrophys. Obs., **58**, 113, 1985.